\scrollmode
\documentclass[doublespacing]{elsart}

\usepackage{epsfig}

\usepackage{amssymb}
\usepackage{amsfonts}

\include{epsf}
\usepackage{latexsym}
\usepackage{indentfirst}
\usepackage{multicol}
\usepackage{array}
\usepackage{exscale}

\begin{document}

\begin{frontmatter}

% Title, authors and addresses

% use the thanksref command within \title, \author or \address for footnotes;
% use the corauthref command within \author for corresponding author footnotes;
% use the ead command for the email address,
% and the form \ead[url] for the home page:
% \title{Title\thanksref{label1}}
% \thanks[label1]{}
% \author{Name\corauthref{cor1}\thanksref{label2}}
% \ead{email address}
% \ead[url]{home page}
% \thanks[label2]{}
% \corauth[cor1]{}
% \address{Address\thanksref{label3}}
% \thanks[label3]{}

\title{Butterfly Hysteresis and Slow Relaxation of the Magnetization in
(Et$_4$N)$_3$Fe$_2$F$_9$: Manifestations of a Single-Molecule Magnet}

% use optional labels to link authors explicitly to addresses:
% \author[label1,label2]{}
% \address[label1]{}
% \address[label2]{}

\author[Bern]{Ralph Schenker},
\author[Basel]{Michael N. Leuenberger},
\author[Bern]{Gr\'egory Chaboussant},
\author[Bern]{Hans U. G\"{u}del\corauthref{cau}},
\author[Basel]{and Daniel Loss}
\corauth[cau]{Corresponding author, hans-ulrich.guedel@iac.unibe.ch, fax: +41 31
631 43 99.}

\address[Bern]{Departement f\"{u}r Chemie und Biochemie,
Universit\"{a}t Bern,\\Freiestrasse 3, CH-3000 Bern 9, Switzerland}
\address[Basel]{Departement f\"{u}r Physik und Astronomie, Universit\"{a}t Basel,\\
 Klingelbergstrasse 82, 4056 Basel, Switzerland}

\begin{abstract}
(Et$_4$N)$_3$Fe$_2$F$_9$ exhibits a butterfly--shaped hysteresis below 5 K when
the magnetic field is parallel to the threefold axis, in accordance with a very
slow magnetization relaxation in the timescale of minutes. This is attributed to
an energy barrier $\Delta=2.40$ K resulting from the $S=5$ dimer ground state of
{[Fe$_2$F$_9$]}$^{3-}$ and a negative axial anisotropy. The relaxation partly
occurs via thermally assisted quantum tunneling. These features of a
single-molecule magnet are observable at temperatures comparable to the barrier
height, due to an extremely inefficient energy exchange between the spin system
and the phonons. The butterfly shape of the hysteresis arises from a phonon
avalanche effect.
%To our knowledge the {[Fe$_2$F$_9$]}$^{3-}$ dimer is the
%smallest spin cluster so far to exhibit slow relaxation.

\end{abstract}

\end{frontmatter}

\section{Introduction}
The discovery of macroscopic quantum spin tunneling and molecular magnetic
hysteresis at cryogenic temperatures in the spin cluster compounds
[Mn$_{12}$O$_{12}$(OAc)$_{16}$(H$_{2}$O)$_{4}$] $\cdot$2HOAc$\cdot$4H$_{4}$O ($\bf
Mn_{12}$),\cite{Caneschi1991,Villain1994,Thomas1996,Friedman1996}
[Fe$_{8}$O$_{2}$(OH)$_{12}$(tacn)$_{6}$]$^{8+}$ (tacn = 1,4,7--triazacyclononane)
($\bf Fe_8$),\cite{Sangregorio1997} and various Mn$_4$ clusters [Mn$_4$O$_3$X
(OAc)$_3$(dbm)$3$] (dmb$^{-}$ = dibenzoylmethanate, X = F, Cl, Br, OAc) ($\bf
Mn_4$)\cite{Aubin1998,Andres2000} has revitalized the field of molecular
magnetism. Chemists have joined forces with physicists in an attempt to fully
characterize and understand the new phenomena and to develop strategies to design
and prepare other members of this new family of single molecule magnets (SMMs). So
far a number of further SMMs have been discovered, mostly by means of ac
susceptibility, among them several manganese
clusters,\cite{Aubin1996,Aubin1999,Yoo2000,Aubin2001} as well as tetranuclear
vanadium\cite{Castro1998} and iron\cite{Barra1999} complexes. All these compounds
contain clusters with at least four paramagnetic centers.
\par
Here we report low-temperature magnetic properties of
(Et$_{4}$N)$_{3}$Fe$_{2}$F$_{9}$ ($\bf Fe_2$), which clearly show effects of slow
relaxation and quantum tunneling. To our knowledge the [Fe$_{2}$F$_{9}$]$^{3-}$
dimer is the smallest exchange-coupled unit so far to exhibit this type of
phenomenon.

\section{Experimental}
\begin{figure}[!h]
\epsfxsize=8.5cm \center\epsffile{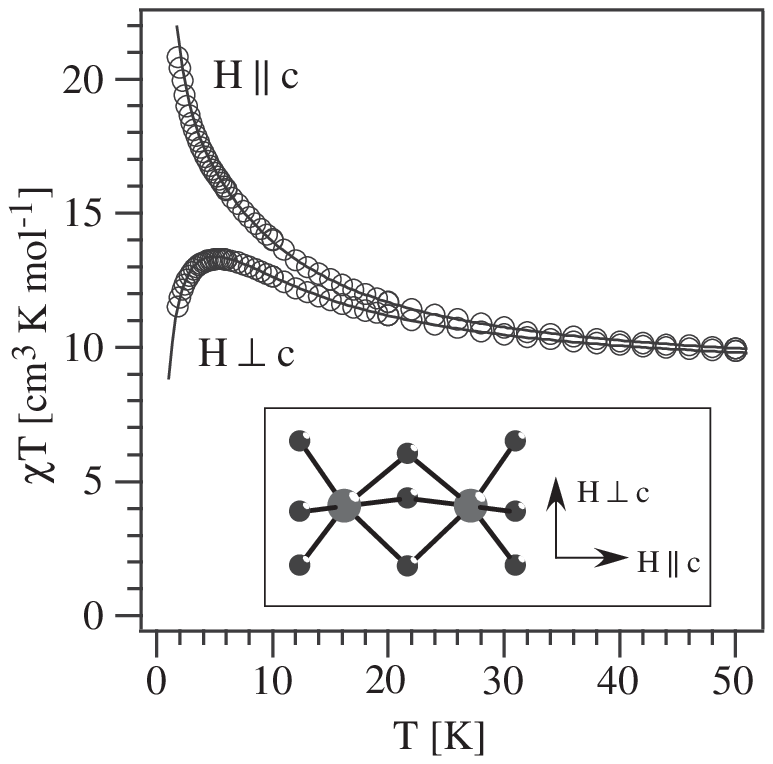}
\caption{}\label{susceptibility}
\end{figure}

(Et$_{4}$N)$_{3}$Fe$_{2}$F$_{9}$ was prepared as small needle-shaped crystals
according to ref.~\cite{SchenkerIC2001}. The product was characterized by X-ray
powder diffraction using the structural information from
ref.~\cite{SchenkerZAAC2001}. It crystallizes in the space group P6$_3$/m with the
trigonal axes of the [Fe$_{2}$F$_{9}$]$^{3-}$ dimers parallel to the hexagonal $c$
axis, see the inset of Fig.~\ref{susceptibility}. Due to their water sensitivity,
single crystals of $\bf Fe_2$ were sealed in glass capillaries for the magnetic
measurements. The crystals were perfectly transparent, indicating that no
hydrolysis occurred.
\par
Magnetic measurements were performed using a Superconducting Quantum Interference
Device (SQUID) magnetometer (Quantum Design MPMS-XL-5 with a 5 T (50 kG) magnet),
with the magnetic field $H$ parallel and perpendicular to the crystallographic $c$
axis.

\section{Results}
The single-crystal magnetic susceptibility is shown for $H \parallel c$ and $H
\perp c$ orientations in Fig.~\ref{susceptibility}. Below 25 K it is highly
anisotropic, and was fitted using the spin Hamiltonian
\begin{equation}
\mathcal{H}=J({\hat S}_1\cdot{\hat S}_2)+D\sum\limits_{i=1,2}
\left[{\hat S}_{z_i}^2-\frac{1}{3}(S_i(S_i+1)\right]
+\sum\limits_{i=1,2}g_i\mu_B{\hat S_i}\cdot H
\end{equation}
with $J=-2.23$ K, $D=-0.215$ K, and $g$ = 2.00. The exchange splitting is
ferromagnetic with an $S$ = 5 dimer ground state, separated by 11.2 K from the
next higher $S$ = 4 state. The single-ion axial anisotropy parameter $D$ can be
correlated with the zero-field-splitting (ZFS) parameter of the dimer ground state
by $D_{S=5}=4/9\cdot D=-0.096$ K.\cite{Kahn1993}
\par
\begin{figure}[!h]
\epsfxsize=8.5cm \center\epsffile{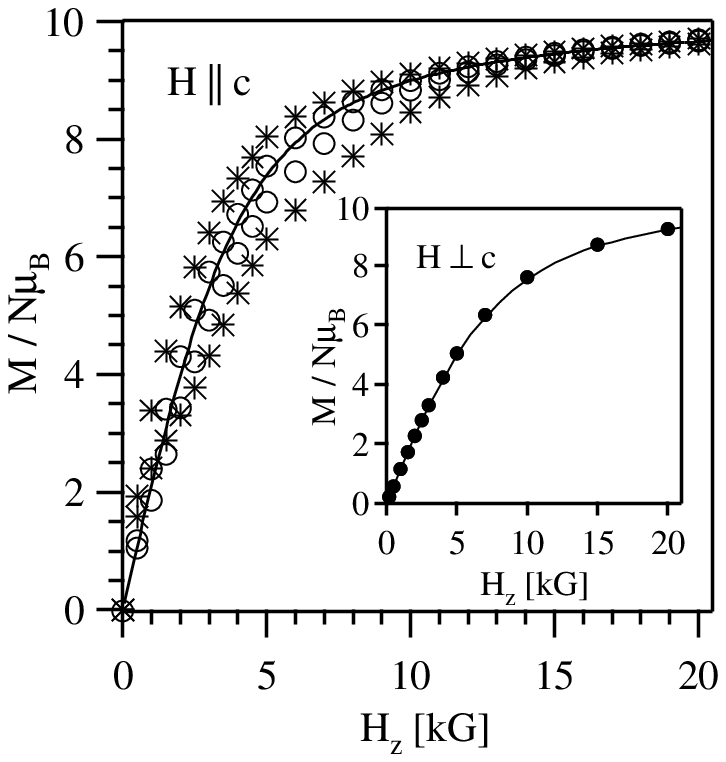} \caption{}\label{hysteresis}
\end{figure}
\begin{figure}[!h]
\epsfxsize=8.5cm \center\epsffile{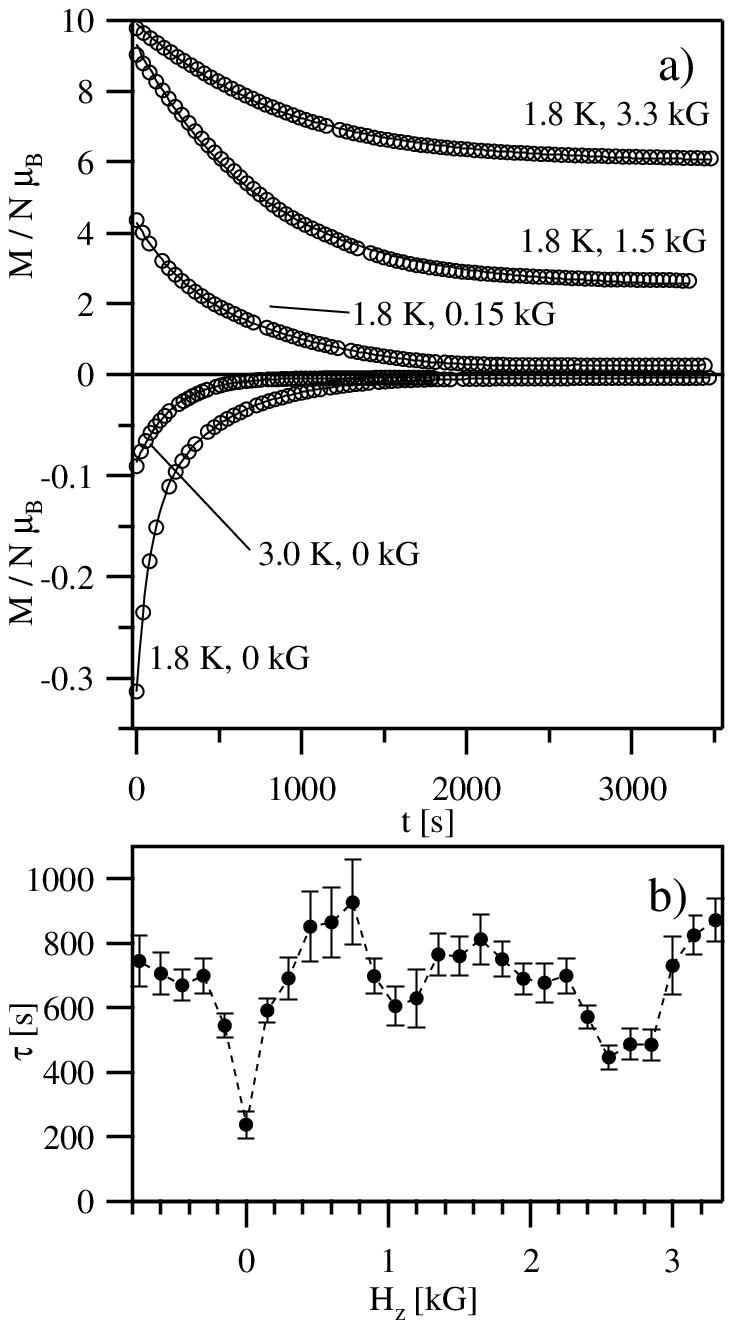} \caption{}\label{relaxation}
\end{figure}

Magnetization measurements with $H \perp c$ showed the expected behavior for the
above Hamiltonian and parameters, see the inset of Fig.~\ref{hysteresis}. The
results of magnetization measurements with $H \parallel c$ ($H_z$) were unusual.
Fig.~\ref{hysteresis} shows 1.8 K results obtained with two different sweeping
rates $\Gamma$. They exhibit a peculiar butterfly-shaped hysteresis with no
remnant magnetization at $H_z=0$. Similar behavior has been observed in $\bf
Mn_{12}$ at 3.25 K, {\it i.e.} just above the blocking
temperature.\cite{Paulsen1995} The breadth of the wings increases with decreasing
temperature and with increasing $\Gamma$. This is direct evidence for a slow
relaxation process at 1.8 K. The effect is strongly $T$ dependent, and we could
observe it up to 5 K. For an infinitely slow $\Gamma$ the butterfly hysteresis
would disappear and the adiabatic curve (full line in Fig.~\ref{hysteresis}) would
be obtained. In order to quantitatively determine the relaxation dynamics, we
performed magnetization relaxation experiments at fixed magnetic field $H_z$ and
temperature $T$, after saturation at 30 kG and subsequent quick reduction
($\Gamma\approx150$ G/s) to the indicated $H_z$ value. Typical decay curves are
depicted in Fig.~\ref{relaxation}a for some selected values of $H_z$ and $T$. The
relaxation depends on both. At 1.8 K and $H_z$ = 3.3 kG the relaxation is so slow
that the equilibrium magnetization has not been reached after 40 minutes. For
$H_z=0$ at 1.8 K, on the other hand, the decay is about 4 times faster. At all
fields the relaxation rate increases with temperature, as shown for $T=1.8$ K and
3.0 K at $H_z=0$ in the lower panel of Fig.~\ref{relaxation}a. Relaxation times
$\tau$ were obtained from single-exponential fits to the relaxation curves for
various $H_z$ and $T$ values. The field dependence of $\tau$ at 1.8 K, shown in
Fig.~\ref{relaxation}b, exhibits 3 distinctive dips at $H_z=0$, 1.1, and 2.7 kG.
The temperature dependence of $\tau$ for a given value of $H_z$ was fitted with
the Arrhenius law
\begin{equation}
\tau=\tau_0\cdot exp(\Delta E/kT)\quad .
\end{equation}
At $H_z=0$ the obtained parameter values are $\tau_0=70$ s and $\Delta E=2.2$ K,
corresponding to $\tau=238$ s at 1.8 K.
\par
\begin{figure}[!h]
\epsfxsize=8.5cm\center\epsffile{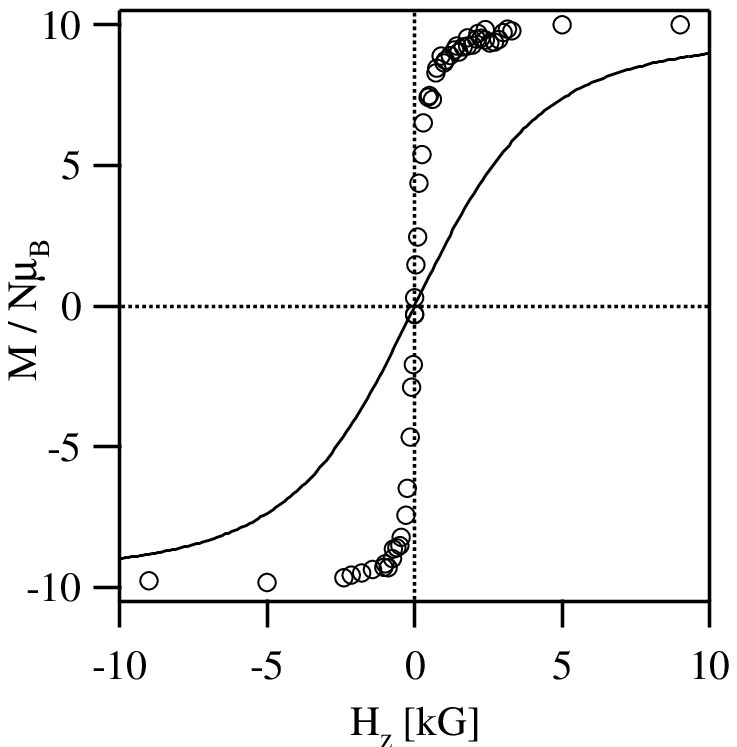} \caption{}\label{avalanche}
\end{figure}

Fig.~\ref{avalanche} shows magnetization data obtained as follows: After
saturation at +30 kG the field was quickly ($\Gamma\approx150$ G/s) reduced to the
given $H_z$ values, including negative ones. The magnetization was measured within
50 s after reaching $H_z$. During such a fast sweep from +30 kG to $-$2.5 kG, {\it
e.g.}, the magnetization switches instantaneously from positive to negative
saturation when crossing $H_z=0$.

\section{Analysis and Discussion}
Except for the butterfly hysteresis, the low-temperature magnetization curves in
Fig.~\ref{hysteresis} are perfectly reproduced with the parameter values derived
from the DC magnetic susceptibility data in the temperature range 1.8 K -- 300 K.
Furthermore, the shortest Fe--Fe distances  between neighboring dimers are at
least 8 {\AA},\cite{SchenkerZAAC2001} which makes interdimer interactions
extremely inefficient. Therefore the observed slow relaxation and butterfly
hysteresis must be a genuine property of the [Fe$_{2}$F$_{9}$]$^{3-}$ dimer.
\par
\begin{figure}[!h]
\epsfxsize=8.5cm\center\epsffile{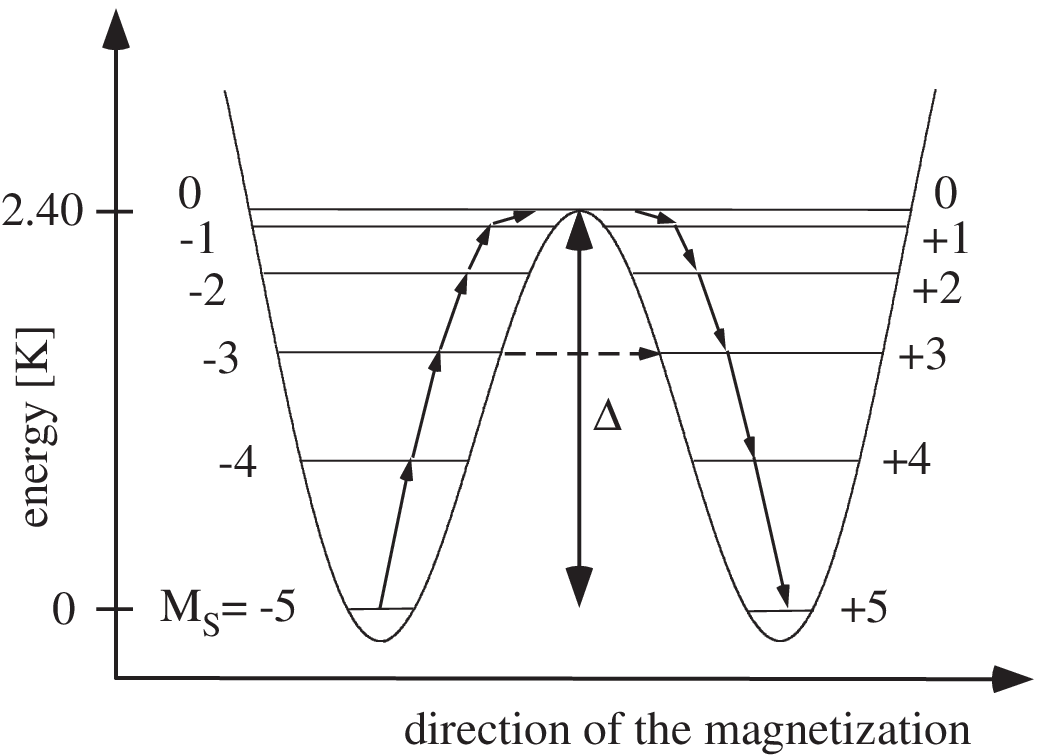} \caption{}\label{potential}
\end{figure}

Our interpretation is based on a comparison with the observed behavior of
single-molecule magnets such as $\bf Mn_{12}$\cite{Caneschi1991,Thomas1996}, as
well as the formalism developed for their interpretation. The appropriate picture
representing the ZFS of the $S=5$ dimer ground state with the negative $D_{S=5}$
is shown in Fig.~\ref{potential}. The barrier height in Fig.~\ref{potential} at
$H_z=0$ is $\Delta=(|D_{S=5}|)S^{2}=2.40$ K. This is slightly larger than the
kinetic activation energy $\Delta E=2.2$ K determined from the magnetization
relaxation. This difference is small, an indication that, if tunneling processes
do contribute to the relaxation at the lowest temperatures, this contribution is
not a dominant one.  In the $C_{3h}$ dimer symmetry, the full anisotropy
Hamiltonian includes the higher--order terms $B_{4}^{0}\hat{O}_{4}^{0}$, $
B_{6}^{0}\hat{O}_{6}^{0}$, and $B_{6}^{6}\hat{O}_{6}^{6}$.\cite{Abragam1970?1} The
$B_{6}^{6}\hat{O}_{6}^{6}$ term mixes wavefunctions with $\Delta M_S=\pm6$. This
allows for resonant tunneling at $H_z=0$ between the $M_S$ levels $-3/+3$ and at
applied fields when the $M_S=-2/+4$ and $-1/+5$ levels cross, leading to
significantly lower values of $\tau$ at these fields and hence to dips in a $\tau$
vs $H_z$ plot. Therefore the observed dips in Fig.~\ref{relaxation}b are direct
evidence for tunneling processes in $\bf Fe_2$. The dips are located at $H_z=0$,
1.1, and 2.7 kG instead of 0, 1.4, and 2.85 kG, respectively, calculated without
the higher-order terms. This indicates nonzero values for the parameters $B_4^0$
and $B_6^0$. In $\bf Fe_2$, the drop of $\tau$ due to resonant tunneling compared
to pure thermal relaxation at non-resonant fields is only about 25-75\%, see
Fig.~\ref{relaxation}b. In contrast, $\tau$ decreases by 1--2 orders of magnitude
in $\bf Mn_{12}$ when resonant tunneling is possible.\cite{Thomas1996} This is in
line with our conclusion from the kinetic energy barrier, that although they
occur, quantum tunneling processes are not dominant in $\bf Fe_2$. Instead, the
purely thermal relaxation is dominant at all temperatures and field values. This
distinctly different behavior from other SMMs is due to the combination of two
facts: i) the thermal barrier in $\bf Fe_2$ does not exceed the barrier for
tunneling processes by more than a factor of 1.6; and ii) the tunneling is phonon
assisted and thus slowed down by an extremely slow energy exchange between the
spin system and the phonon bath. This latter effect will now be analyzed.
\par
The prefactor $\tau_0=70$ s obtained from the Arrhenius analysis of the relaxation
dynamics at $H_z=0$ is unusually large, $7-11$ orders of magnitude larger than in
$\bf Mn_{12}$\cite{Caneschi1991,Villain1994,Thomas1996,Friedman1996} and other
SMMs~\cite{Sangregorio1997,Aubin1998,Aubin1996,Aubin1999,Yoo2000,Aubin2001,Castro1998,Barra1999}.
It is much larger than the timescale of ac magnetic susceptibility measurements,
which are usually used to characterize and quantify the relaxation behavior in
SMMs. Only due to this extremely high $\tau_0$ value are we able to observe a slow
relaxation phenomenon below 5 K with a barrier height of only $\Delta=2.40$ K. In
the SMMs reported so far, for comparison, $\Delta$ ranges from 7.6 K in
[Mn$_4$O$_3$F(OAc)$_3$(dbm)$_3$]\cite{Andres2000} to 79 K in
(PPh$_4$)[Mn$_{12}$O$_{12}$(O$_2$CEt)$_{16}$(H$_2$O)$_4$]\cite{Aubin1999}. As
indicated in Fig.~\ref{potential}, the magnetization in {\bf Fe$_2$} can relax
both by thermal activation over the energy barrier (full arrows) and by tunneling
between the $M_S=-3/+3$ levels (dashed arrows). Both mechanisms rely on an
interaction between the spin system and the crystal lattice degrees of freedom,
{\it i.e.} on spin-phonon coupling. In this process, energy quanta corresponding
to the energy differences between adjacent spin levels are exchanged between the
spin system and the phonon bath. The extremely large $\tau_0$ in $\bf Fe_2$
results from an extreme inefficiency of these processes below 5 K, and this can be
ascribed to the following two factors: a small spin-phonon coupling parameter
$g_0$ and a small density of phonon states in resonance with the energy spacings
in Fig.~\ref{potential}. Quantitatively, the rate constant of a $\Delta M_S= \pm1$
transition induced by phonon absorption or emission can be expressed
as:\cite{Leuenberger2000}
\begin{equation}
W_{M_S\pm 1,M_S}=\frac{g_0^2S_{\pm 1}}{48\pi\rho v^5\hbar^4}
\frac{(\varepsilon_{M_S\pm 1}-\varepsilon_{M_S})^3} {e^{(\varepsilon_{M_S\pm
1}-\varepsilon_{M_S})/kT}-1}\, ,
\end{equation}
where $S_{\pm1}=(S\mp M_S)(S\pm M_{S+1})(2M_S\pm1)^2$. The number of phonons
involved in the process depends on their energy ($\varepsilon_{M_S\pm
1}-\varepsilon_{M_S}$) and the energy distribution of the density of states,
expressed by the sound velocity $v$ in eq. 3, the mass density $\rho=1.36$
g/cm$^{3}$\cite{SchenkerZAAC2001} and the Bose population factor
($e^{(\varepsilon_{M_S\pm 1}-\varepsilon_{M_S})/kT}-1$).\cite{Leuenberger2000} We
note that ($\varepsilon_{M_S\pm 1}-\varepsilon_{M_S}$) is about 1 order of
magnitude smaller in $\bf Fe_2$ than in $\bf Mn_{12}$ and other SMMs. In all the
SMMs reported so
far,\cite{Caneschi1991,Villain1994,Thomas1996,Friedman1996,Sangregorio1997,Aubin1998,Andres2000,Aubin1996,Aubin1999,Yoo2000,Aubin2001,Castro1998,Barra1999}
the clusters are relatively large in size and contain organic ligands such as
acetate. This gives rise to a large number of vibrational and rotational
low-energy internal modes, leading to a high density of states in the energy range
most relevant for spin-phonon coupling. This is in sharp contrast to $\bf Fe_2$,
in which the [Fe$_2$F$_9$]$^{3-}$ unit consists of atomic ligands with stiff Fe--F
bonds, reflected in the absence of low-energy vibrational or rotational modes.
Compared to the other SMMs, the stiffness of the [Fe$_2$F$_9$]$^{3-}$ unit results
in a strongly increased average sound velocity $v$ in the crystal. Thus in $\bf
Fe_2$ the density of phonon states $\rho_{ph}\propto 1/v^{3}$\cite{Abragam1970?2}
is extremely low in the relevant energy range. The magnitude of the spin-phonon
coupling parameter $g_0$ is governed by the strengths of spin-orbit and
electron-phonon (orbit-lattice) couplings.\cite{VanVleck1940,Mattuck1960} The
latter is expected to be smaller in $\bf Fe_2$ compared to other SMMs due to the
great stiffness of the [Fe$_2$F$_9$]$^{3-}$ unit. As spin-orbit coupling - in
terms of the ZFS parameter $D_S$ for the ground spin state - is also smaller in
$\bf Fe_2$ compared to other
SMMs,\cite{Caneschi1991,Villain1994,Thomas1996,Friedman1996,Sangregorio1997,Aubin1998,Andres2000,Aubin1996,Aubin1999,Yoo2000,Aubin2001,Castro1998,Barra1999}
we can expect a smaller spin-phonon coupling constant $g_0$ in $\bf Fe_2$. Thus
the efficiency of spin-phonon coupling is greatly reduced in $\bf Fe_2$ compared
to other SMMs, and the smaller ($\varepsilon_{M_S\pm 1}-\varepsilon_{M_S}$)$^{3}$,
the larger $v^{5}$, and the smaller $g_0^{2}$ factors all contribute to the
drastic reduction of the transition rate constant $W$ in $\bf Fe_2$. As a
consequence, the prefactor $\tau_0$ in the Arrhenius expression is increased by
7--11 orders of magnitude.
\par
The magnetization data shown in Fig.~\ref{avalanche} obtained with
$\Gamma\approx150$ G/s exhibit a very fast reversal of the magnetization around
$H_z=0$. We ascribe this to a so-called phonon avalanche,\cite{Abragam1970?2}
arising from a resonance of phonon absorption on one side of the barrier with
phonon emission on the other side. A similar phenomenon has been observed in $\bf
Mn_{12}$ at $H_z\neq0$, induced by heat pulses\cite{Paulsen1995} or extremely high
sweeping rates.\cite{Hernandez1999} The butterfly shape of the hysteresis in $\bf
Fe_{2}$ is very similar to that reported for
K$_6$[V$_{15}$As$_6$O$_{42}$(H$_2$O)]\ $\cdot$ 8H$_2$O,\cite{Chiorescu2000} but in
the latter system it has been attributed to a phonon bottleneck arising from
dissipative spin reversal in the absence of an energy barrier. Our results show
that a butterfly-shaped hysteresis can also occur in the presence of a barrier.
\par
Due to the phonon avalanche in $\bf Fe_2$, the {\it reversal} of the magnetization
at $H_z=0$ is a much faster process than its {\it relaxation} to equilibrium. The
smaller the sweeping rate $\Gamma$ in the magnetization experiment, the closer is
the system to equilibrium, as is nicely demonstrated by the decrease of the
butterfly hysteresis with $\Gamma$ in Fig.~\ref{hysteresis}. After a fast
($\Gamma\approx150$ G/s) reduction of $H_z$ to 0, the system is therefore still
out of equilibrium at the beginning of the relaxation measurement. Since its
reversal is so fast, the magnetization then overshoots, i.e. it becomes slightly
negative, before it relaxes to its adiabatic value. Therefore there is an
observable negative magnetization relaxation at $H_z=0$, see
Fig.~\ref{relaxation}a, despite the fact that the butterfly hysteresis passes
through the origin of the coordinates in Fig.~\ref{hysteresis}.
\par
In conclusion, we have shown $\bf Fe_2$ to exhibit unusually slow magnetization
relaxation effects below 5 K. These can be ascribed to an energy barrier
$\Delta=2.40$ K resulting from the $S=5$ dimer ground state of
[Fe$_2$F$_9$]$^{3-}$ and a negative axial anisotropy. The relaxation partly occurs
via a thermally-assisted quantum tunneling process. Compared to other SMMs, the
energy exchange between the spin system and the phonons is extremely inefficient
in $\bf Fe_2$. Therefore the SMM phenomenon is observed at temperatures comparable
to the barrier height and far above the blocking temperature, which for $\bf Fe_2$
is expected to lie in the mK regime. To our knowledge the $\bf Fe_2$ dimer is the
smallest spin cluster so far to exhibit slow relaxation.
\par
{\bf Acknowledgement.} This work has been supported by the Swiss National Science
Foundation and by the TMR program Molnanomag of the EU (HPRN-CT-1999-00012).

%\pagebreak
\section*{Figure Captions}

{\bf Fig.~\ref{susceptibility}} Temperature dependence of the magnetic
susceptibility $\chi T$ of a single crystal of (Et$_4$N)$_3$Fe$_2$F$_9$ in the
range 1.8--50 K (circles). The crystal was oriented with its $c$ axis parallel
($H\parallel c$) and perpendicular ($H\perp c$) to the magnetic field of 1 kG. The
solid lines are calculated using eq. 1 with the parameters $J=-2.23$ K, $D=-0.215$
K, and $g$ = 2.00. The inset shows the [Fe$_{2}$F$_{9}$]$^{3-}$ dimer molecule and
the magnetic field direction for both $H\parallel c$ and $H\perp c$ orientations.

{\bf Fig.~\ref{hysteresis}} Positive wings of the butterfly hystereses of the
magnetization in $H \parallel c$ orientation at 1.8 K for sweeping rates of
$\Gamma\approx$3 G/s ($\ast$) and $\Gamma\approx$1 G/s ($\circ$). The 1.8 K
magnetization in $H \perp c$ is shown in the inset. The solid lines are the
adiabatic curves, calculated with $J=-2.23$ K, $D=-0.215$ K, and $g$ = 2.00
derived from susceptibility measurements.

{\bf Fig.~\ref{relaxation}} a) Relaxation of the magnetization at fixed $H_z$ and
$T$ as indicated, after saturation at 30 kG and quick ($\Gamma\approx150$ G/s)
sweep of the field to the indicated $H_z$ value. The solid lines are fits to a
single exponential law. Note the different scale for the $H_z=0$ measurements. b)
Dependence of the relaxation time $\tau$ from $H_z=0$ with errors (3$\sigma$). The
dotted line is a guide to the eyes.

{\bf Fig.~\ref{avalanche}} Magnetization data for $H \parallel c$ orientation at
1.8 K obtained as follows: After saturation at +30 kG, the field was quickly
($\Gamma\approx150$ G/s) reduced to the given $H_z$ values including negative
ones, and the magnetization was measured within 50 s after reaching $H_z$. The
solid line is the calculated adiabatic magnetization curve at 1.8 K, see also
Fig.~\ref{hysteresis}.

{\bf Fig.~\ref{potential}} Schematic plot of the potential energy {\it vs}
magnetization direction for a $S=5$ ground state split by a negative axial
zero--field splitting at $H_z=0$. The potential energy barrier is
$\Delta=(|D_{S=5}|)S^{2}=2.40$ K. The full and broken arrows denote the thermal
relaxation and possible tunneling pathways, respectively.

\pagebreak
\section*{Figures}
%%%%%%%%%%%%%%%%%%%%%%%%%%%%%%%%%%%%%%%%%%%%%%%%%%%%%%%%%%%%%%%%%%%%%%%%%%%%%%%%%%%%%%%%%%%%%%%%%%%%%%%%%%%%%%%%%%%%%
%\vspace*{5cm}\begin{center}{\bf Figure 1}\end{center}
%%%%%%%%%%%%%%%%%%%%%%%%%%%%%%%%%%%%%%%%%%%%%%%%%%%%%%%%%%%%%%%%%%%%%%%%%%%%%%%%%%%%%%%%%%%%%%%%%%%%%%%%%%%%%%%%%%%%%

\pagebreak
%%%%%%%%%%%%%%%%%%%%%%%%%%%%%%%%%%%%%%%%%%%%%%%%%%%%%%%%%%%%%%%%%%%%%%%%%%%%%%%%%%%%%%%%%%%%%%%%%%%%%%%%%%%%%%%%%%%%%
%\vspace*{5cm}\begin{center}{\bf Figure 2}\end{center}
%%%%%%%%%%%%%%%%%%%%%%%%%%%%%%%%%%%%%%%%%%%%%%%%%%%%%%%%%%%%%%%%%%%%%%%%%%%%%%%%%%%%%%%%%%%%%%%%%%%%%%%%%%%%%%%%%%%%%

\pagebreak
%%%%%%%%%%%%%%%%%%%%%%%%%%%%%%%%%%%%%%%%%%%%%%%%%%%%%%%%%%%%%%%%%%%%%%%%%%%%%%%%%%%%%%%%%%%%%%%%%%%%%%%%%%%%%%%%%%%%%
%\vspace*{5cm}\begin{center}{\bf Figure 3}\end{center}
%%%%%%%%%%%%%%%%%%%%%%%%%%%%%%%%%%%%%%%%%%%%%%%%%%%%%%%%%%%%%%%%%%%%%%%%%%%%%%%%%%%%%%%%%%%%%%%%%%%%%%%%%%%%%%%%%%%%%

\pagebreak
%%%%%%%%%%%%%%%%%%%%%%%%%%%%%%%%%%%%%%%%%%%%%%%%%%%%%%%%%%%%%%%%%%%%%%%%%%%%%%%%%%%%%%%%%%%%%%%%%%%%%%%%%%%%%%%%%%%%%%
%\vspace*{5cm}\begin{center}{\bf Figure 4}\end{center}
%%%%%%%%%%%%%%%%%%%%%%%%%%%%%%%%%%%%%%%%%%%%%%%%%%%%%%%%%%%%%%%%%%%%%%%%%%%%%%%%%%%%%%%%%%%%%%%%%%%%%%%%%%%%%%%%%%%%%%%%

\pagebreak
%%%%%%%%%%%%%%%%%%%%%%%%%%%%%%%%%%%%%%%%%%%%%%%%%%%%%%%%%%%%%%%%%%%%%%%%%%%%%%%%%%%%%%%%%%%%%%%%%%%%%%%%%%%%%%%%%%%%%%
%\vspace*{5cm}\begin{center}{\bf Figure 5}\end{center}
%%%%%%%%%%%%%%%%%%%%%%%%%%%%%%%%%%%%%%%%%%%%%%%%%%%%%%%%%%%%%%%%%%%%%%%%%%%%%%%%%%%%%%%%%%%%%%%%%%%%%%%%%%%%%%%%%%%%%%%%


\begin{thebibliography}{00}
% \bibitem{label}
% Text of bibliographic item

% notes:
% \bibitem{label} \note

% subbibitems:
% \begin{subbibitems}{label}
% \bibitem{label1}
% \bibitem{label2}
% If there is a note, it should come last:
% \bibitem{label3} \note
% \end{subbibitems}

%Mn12 acetate
% 1 Caneschi
\bibitem{Caneschi1991} A. Caneschi, D. Gatteschi, R. Sessoli, A.--L. Barra, L. C. Brunel, and M.
Guillot, J. Am. Chem. Soc. {\bf 113}, 5873 (1991).
% 2 Villain
\bibitem{Villain1994} J. Villain, F. Hartman--Boutron, R. Sessoli, A. Rettori,Europhys. Lett. {\bf 27}, 159 (1994).
% 12 Fort
%\bibitem{Fort1998} A. Fort, A. Rettori, J. Villain, D. Gatteschi, R. Sessoli, Phys. Rev. Lett. {\bf 80}, 612 (1998).
% 3 Relaxation (ii)
\bibitem{Thomas1996} L. Thomas, F. Lionti, R. Ballou, D. Gatteschi, R. Sessoli, B. Barbara,
Nature {\bf 12}, 145 (1996).
% 4 Friedman
\bibitem{Friedman1996} J. R. Friedman, M. P. Sarachik, J. Tejada, R. Ziolo, Phys. Rev.
Lett. {\bf 76}, 3830 (1996).
% 5 Fe8, hysteresis
\bibitem{Sangregorio1997} C. Sangregorio, T. Ohm, C. Paulsen, R. Sessoli, D.
Gatteschi, Phys. Rev. Lett {\bf 78}, 4645 (1997).
% 6 Mn4, hysteresis
\bibitem{Aubin1998} S. M. J. Aubin, N. R. Dilley, L. Pardi, J. Krzystek, M. W. Wemple,
L.-C. Brunel, M. Brian Maple, G. Christou, D. N. Hendrickson, J. Am. Chem. Soc.
{\bf 120}, 4991 (1998).
% 7 Mn4, INS
\bibitem{Andres2000}H. P. Andres, R. Basler, H. U. G\"{u}del, G. Aromi, G.
Christou, H. B\"{u}ttner, B. Ruffl\'{e}, J. Am. Chem. Soc. {\bf 122}, 12469
(2000).
% 7 Mn4 cubane
\bibitem{Aubin1996} S. M. L. Aubin, M. W. Wemple, D. M. Adams, H.--L. Tsai, G. Christou, D. N.
Hendrickson, J. Am. Chem. Soc. {\bf 118}, 7746 (1996).
% 8 Mn12 ion Ph
\bibitem{Aubin1999} S. M. J. Aubin, Z. Sun, L. Pardi, J. Krzystek,  K. Folting,
L.--C. Brunel, A. Rheingold, G. Christou, D. N. Hendrickson, Inorg. Chem. {\bf
38}, 5329 (1999).
% 9 Mn2Mn3 planar
\bibitem{Yoo2000} J. Yoo, E. K. Brechin, A. Yamaguchi, M. Nakano, J. C. Huffman, A. L. Maniero,
L.--C. Brunel, K. Awaga, H. Ishimoto, G. Christou, D. N. Hendrickson, Inorg. Chem.
{\bf 39}, 3615 (2000).
% 10 Mn12 Jahn-Teller
\bibitem{Aubin2001} S. M. J. Aubin, Z. Sun, H. J. Eppley, E. M. Rumberger, I. A.
Guzei, K. Folting, P. K. Gantzel, A. L. Rheingold, G. Christou, D. N. Hendrickson,
Inorg. Chem. {\bf 40}, 2127 (2001).
% 11 V4 butterfly
\bibitem{Castro1998} S. L. Castro, Z. Sun, C. M. Grant, J. C. Bollinger,
D. N. Hendrickson, G. Christou, J. Am. Chem. Soc. {\bf 120}, 2365 (1998).
% 12 Fe4
\bibitem{Barra1999} A.--L. Barra, A. Caneschi, A. Cornia, F. Fabrizi de Biani, D. Gatteschi,
C. Sangregoria, R. Sessoli, L. Sorace, J. Am. Chem. Soc. {\bf 121}, 5302 (1999).
% 13 Synthesis
\bibitem{SchenkerIC2001} R. Schenker, H. Weihe, H. U. G\"{u}del, Inorg. Chem. {\bf 40}, 4319 (2001).
% 14 Structure
\bibitem{SchenkerZAAC2001} K. W. Kr\"{a}mer, R. Schenker, J. Hauser, H. Weihe,
H. U. G\"{u}del, H.-B. B\"{u}rgi, Z. Anorg. Allg. Chem. {\bf 627}, 2511 (2001).
% Kahn book
\bibitem{Kahn1993} O. Kahn, {\it Molecular Magnetism}; VCH Publishers: New York,
1993, 142.
% 15 Paulsen1995 avalanche
\bibitem{Paulsen1995} C. Paulsen, J.--G. Park, B. Barbara, R. Sessoli, A. Caneschi,
J. Magn. Magn. Mater. {\bf 140}, 1891 (1995).
% 16 Abragam&Bleney
\bibitem{Abragam1970?1} A. Abragam, B. Bleaney, {\it Electron Paramagnetic Resonance of Transition Metal
Ions}; Clarendon Press: Oxford, 1970, 338.
% 17 Leuenberger
\bibitem{Leuenberger2000} M. N. Leuenberger, D. Loss, Phys. Rev. B {\bf 61}, 1286 (2000).
% 18 Black Cr2Cl9 NCA
%\bibitem{Black1975} for a normal coordinate analysis on the related
%[Cr$_2$Cl$_9$]$^{3-}$ ion, see: J. D. Black, J. T. R. Dunsmuir, I. W. Forrest, A.
%P. Lane, Inorg. Chem. {\it 14}, 1257 (1975).
% 19 Abragam&Bleney
\bibitem{Abragam1970?2} A. Abragam, B. Bleaney, {\it Electron Paramagnetic Resonance of Transition Metal
Ions}; Clarendon Press: Oxford, 1970, 574.
% 20 vanvleck
\bibitem{VanVleck1940} J. H. Van Vleck, Phys. Rev. {\bf 57}, 426 (1940).
% 21 Mattuck
\bibitem{Mattuck1960} R. D. Mattuck, M. W. P. Strandberg, Phys. Rev. {\it 119},
1204 (1960).
% 19 avalanche2
\bibitem{Hernandez1999} E. del Barco, J. M. Hernandez, M. Sales, J. Tejada, H. Rakoto, J. M. Broto, E. M. Chudnovsky,
Phys. Rev. B {\bf 60}, 11898 (1999).
% 20 V15 butterfly
\bibitem{Chiorescu2000} I. Chiorescu, W. Wernsdorfer, A. M\"{u}ller, H. B\"{o}gge, B.
Barbara, Phys. Rev. Lett. {\bf 84}, 3454(2000).
\\ % \vspace*{15cm}
\end{thebibliography}
\end{document}